# Using Few-Shot Learning to Classify Primary Lung Cancer and Other Malignancy with Lung Metastasis in Cytological Imaging via Endobronchial Ultrasound Procedures

Ching-Kai Lin[1,2,3,4], Di-Chun Wei[1]*, Yun-Chien Cheng[1]*,

[1] Department of Mechanical Engineering, College of Engineering, National Yang Ming Chiao Tung University, Hsin-Chu, Taiwan

[2] Department of Medicine, National Taiwan University Cancer Center, Taipei, Taiwan

[3] Department of Internal Medicine, National Taiwan University Hospital, Taipei, Taiwan

[4] Department of Internal Medicine, National Taiwan University Hsin-Chu Hospital, Hsin-Chu, Taiwan

*Corresponding author: dichun1112.en11@nycu.edu.tw, yccheng@nycu.edu.tw

## Abstract

This study develops a computer-aided diagnosis (CAD) system for endobronchial ultrasound (EBUS) procedures to assist physicians in the preliminary identification of lung metastases. The system enables immediate follow-up assessments for metastases in other regions after EBUS, reducing waiting time by more than half and allowing earlier detection and treatment of secondary cancers.In current clinical practice, cytological image data for lung metastases is limited, and morphological similarities among cell types make classification challenging. Existing image-based classification research rarely focuses on lung metastases and lacks solutions tailored to their unique characteristics. Most prior studies rely on publicly available datasets from similar domains, avoiding knowledge transfer challenges. Although Transformer-based models capture global features well, they struggle with local relationships. With limited data, large models are difficult to fine-tune and prone to overfitting.

To address these challenges, this study proposes a few-shot learning classification model. A hybrid pretrained model is used for meta-training, combined with fine-grained image classification and contrastive learning to enhance generalization. Parameter-efficient fine-tuning is then applied using augmented support sets as pseudo-query sets, enabling effective knowledge transfer from meta-training to the target dataset and improving classification. Results show the proposed model, combining broad prior knowledge with efficient fine-tuning, achieved 49.59% accuracy on the test dataset, outperforming other state-of-the-art (SOTA) methods. Precision, recall, and F1 score reached 0.479, 0.496, and 0.488, respectively. Providing more image samples (e.g., 20 images) further improved feature learning, raising accuracy to 55.48% and enhancing discrimination between similar classes. These findings demonstrate the model's potential for recognizing novel or rare categories in clinical scenarios with limited data, opening new possibilities for diagnostic applications.

## 1. Introduction

Cancer is the leading cause of death in Taiwan [1], with the majority of fatal cases associated with cancer metastasis. Early detection and treatment of metastatic cancer are key to reducing mortality. With the increasing availability of imaging technologies such as chest X-rays and computed tomography (CT), identifying pulmonary lesions has become progressively easier. In addition to primary lung cancer, the lungs are also a common site of metastasis for various types of cancer [2], such as breast and colorectal cancers, which frequently spread to the lungs and significantly increase patient mortality risk. When cancer cells spread to the lungs, the condition is referred to as "pulmonary metastatic cancer." There is a substantial difference in cancer staging between primary lung cancer and pulmonary metastasis. Not only are the prognoses markedly different, but the selection of antitumor drugs also varies depending on the tumor type. Therefore, promptly and safely obtaining pulmonary lesion specimens for diagnosis is crucial for subsequent treatment planning.

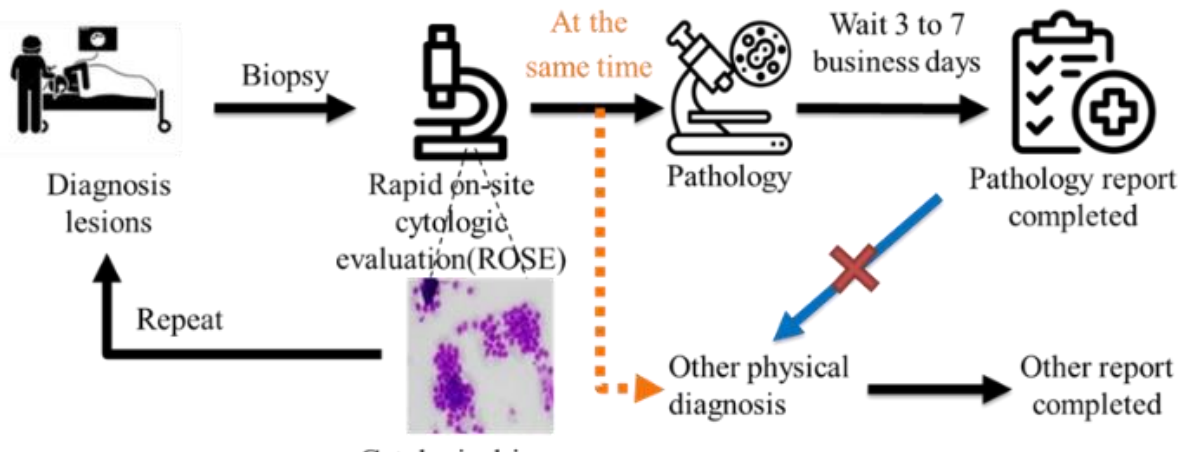

Figure 1. EBUS procedures and diagnosis for lung metastasis

Endobronchial ultrasound (EBUS) is a relatively new minimally invasive technique for examining the lungs. Its procedural workflow is illustrated in Figure 1. Physicians first use different ultrasound probes to identify lesions located in the peripheral lungs or the mediastinum/hilum. Then, biopsy or aspiration procedures are performed to collect tissue samples. During the examination, rapid on-site cytologic evaluation (ROSE) is employed. The freshly obtained specimens are smeared onto slides, rapidly stained, and examined under a microscope to confirm whether the collected samples are adequate, thereby improving diagnostic accuracy [3]. Afterward, the samples are sent to pathology for further analysis, and once the results are available, follow-up treatment plans are arranged. If pulmonary metastasis is diagnosed, additional evaluations and treatments related to metastatic cancer, as indicated by the blue path in Figure 1, are undertaken..

However, current EBUS procedures and the diagnosis of pulmonary metastatic cancer face several challenges:

(1) Long report turnaround time: Pathology results for EBUS specimens typically require 3 to 7 working days. If metastasis confirmation is needed, further tests must be scheduled, leading to a prolonged diagnostic process that delays treatment and increases the risk of further cancer spread.

(2) Limited cytologic imaging data: Clinicians currently rely heavily on patient history and imaging data such as X-rays for diagnosing pulmonary metastasis. Biopsy-confirmed diagnoses are less common, resulting in a scarcity and uneven distribution of cytologic imaging data available for training diagnostic models.

(3) Challenges in cell classification: The lungs are a common metastatic site for many cancers. Due to the morphological similarities among cytologic images from different cancer types, pulmonary metastasis datasets tend to be diverse and complex, making accurate classification more difficult.

To assist clinicians in the preliminary diagnosis of pulmonary metastatic cancer, we have developed a computer-aided diagnostic system. This system can instantly analyze cytologic images following EBUS examination, offering physicians a basis for initial interpretation, as shown by the orange path in Figure 1. By using our system for preliminary classification of cytologic images, clinicians can promptly arrange further metastasis-related examinations and simultaneously plan subsequent treatments. This effectively shortens the waiting time for treatment and achieves the goal of early detection and early intervention.

Given the characteristics of datasets that contain a large number of classes with few samples per class and an imbalanced distribution, previous research has increasingly focused on Few-Shot Learning (FSL) for medical image classification tasks [4–9], achieving promising results. For instance, the team led by Wang [7] applied the WRN28-10 model to classify benign and malignant breast cancer histopathology images and achieved excellent performance. Subsequently, the Mohanta team [8] further advanced multi-class classification for breast cancer pathology images, also demonstrating favorable outcomes.

Other teams, such as Teng et al. [4], focused on classifying lymph node metastases from lung cancer using a deep residual network (ResNet34) and achieved strong classification performance. Medela et al. [5] concentrated on tissue images from various cancer types including colorectal, breast, and lung cancers, using the classical VGG16 convolutional neural network for classification. Their experimental results confirmed the model's robust generalization capabilities. Teh et al. [6] adopted ResNet18 to analyze tissue slides of colorectal and breast cancers, conducting an in-depth analysis of category identification. Xu et al. [9] utilized a lightweight Vision Transformer architecture, DeiT-Tiny, for liver cancer image classification, demonstrating both high inference efficiency and accuracy.

These findings reveal a progression in model choices across the literature. Earlier studies employed commonly used convolutional neural networks such as VGG16 and ResNet34. However, the field has evolved towards deeper and more expressive architectures like WRN28-10, and even toward structures that integrate graph neural networks (GNNs) and self-attention mechanisms, such as GnnNet and DeiT-Tiny. This trend reflects an increasing emphasis on models with strong generalization, scalability, and transferability, enabling more effective few-shot learning performance in medical image classification.

From a data perspective, we also observe that many prior studies utilized similar sources and evaluation strategies in designing medical image classification systems. For example, Teng et al. [4] used a custom-built histopathology dataset containing 1,701 whole slide images (WSIs) for both training and testing, forming a single-source training pipeline. In contrast, Medela et al. [5] adopted a cross-institutional setting, training on a German colorectal cancer dataset (5,000 images) and testing on the BIOEF dataset from Spain, which included images from colorectal adenocarcinoma, breast cancer, and lung cancer (totaling 1,755 images), thereby presenting a classic domain adaptation scenario.

Teh et al. [6] expanded the dataset scale by using KimiaPath24 (23,916 images) as the training set, and the German colorectal dataset (5,000 images) along with PCam (up to 327,680 images) as test sets from the target domain, showcasing a multi-source integration scenario. Wang et al. [7] used MINI-ImageNet as the source dataset and applied it to the medical domain, with testing conducted on the publicly available BreakHis dataset (9,109 images). Mohanta et al. [8] trained on the Lung_Colon dataset (25,000 images) and conducted cross-domain tests on two breast cancer datasets: BreastCancer_IDC_Grades (922 images) and BreakHis (7,909 images). Xu et al. [9] used the large-scale NCT colorectal tissue dataset (100,000 images) for training and tested on the PAIP liver cancer dataset (50 WSIs), showing potential for cross-cancer-type classification.

However, upon reviewing these studies [4–9], we find that most of the proposed methods rely on publicly available datasets for both training and testing, in which each class has a large number of samples and clear morphological distinctions. These conditions differ significantly from the real-world characteristics of our dataset, which involves morphologically similar classes and imbalanced data distribution. Compared to prior studies, the challenges faced in this research are more complex, as our data are derived from real clinical environments, featuring large disparities in sample sizes and similar cellular morphologies across different cancer types. This makes the classification task substantially more difficult. Therefore, directly applying the methods from these studies to our dataset is unlikely to yield ideal performance, highlighting the necessity of developing solutions specifically tailored to the characteristics of real-world clinical data..

Many existing Few-Shot Learning (FSL) methods are developed based on publicly available datasets, such as the MINI-ImageNet dataset, aiming to build a generalizable and transferable model under limited data scenarios. Among them, the team led by Hu [10] proposed a "Pre-train → Meta-learn → Fine-tune" (PMF) training strategy, integrating pre-training, meta-learning, and fine-tuning processes into a streamlined and representative baseline framework. This approach has been subsequently enhanced by later studies. For example, the CAML (Context-Aware Meta-Learning) method proposed by the Fifty team [11], and the SgVA-CLIP (Semantic-Guided Visual Adapting of Vision-Language Models) approach developed by Peng et al. [12], both incorporate language models into the learning pipeline. These methods enable the model to comprehend the semantic meanings of textual labels while learning visual features, thereby improving classification accuracy.

Furthermore, Singh et al. [13] introduced TRIDENT (Transductive Decoupled Variational Inference), which leverages semantic and class latent variables in images to enhance classification performance and generalization under scarce data conditions. Zhou et al. [14] designed LDP-net (Local-global Distillation Prototypical Network), which integrates local and global feature distillation mechanisms to improve the stability and discriminative power of prototype representations in few-shot learning.

From the literature, it is evident that recent advancements in FSL are gradually shifting from purely vision-based approaches to those incorporating category label information [13], and even to the use of pre-trained language models [11–12], such as CLIP (Contrastive Language–Image Pretraining). These multimodal models learn the associations between images and text, enabling the model to understand cross-modal semantics and promoting the development toward multimodal fusion and cross-domain generalization.

However, our dataset exhibits significant domain discrepancies between the training and testing stages, with high morphological similarity among categories in the test set. Although CAML [11] and SgVA-CLIP [12] are recent state-of-the-art methods, literature findings show that pre-trained vision-language models such as CLIP tend to perform poorly when facing large domain shifts and subtle inter-class visual differences. Moreover, in scenarios where upstream and downstream tasks differ considerably, incorporating fine-tuning during the testing stage has been shown to further enhance classification performance [8].

Therefore, this study adopts and refines the "Pre-train → Meta-learn → Fine-tune" (PMF) training strategy proposed by Hu et al. [10], adapting it to better suit the task of classifying cytological images of pulmonary metastatic cancer.

Although existing studies have proposed using fine-tuning to mitigate the performance degradation caused by significant domain discrepancies between training and testing datasets, in the context of few-shot learning, fine-tuning large pre-trained models with extremely limited data is often suboptimal. On one hand, it tends to cause model overfitting; on the other hand, it fails to fully leverage the potential of large models. Moreover, the computational resources required and the number of parameters to be adjusted are relatively high.

To address this challenge, recent research in few-shot learning has introduced a series of Parameter-Efficient Fine-Tuning (PEFT) techniques, which have been explored for use in few-shot scenarios. These methods aim to fine-tune large models using limited data while maintaining efficiency. A summary of relevant PEFT-based studies, which core idea of PEFT is to update only a small subset of the pre-trained model's parameters or to insert lightweight parameter modules, allowing most of the original model weights to remain unchanged. This approach achieves performance comparable to, or even better than, full model fine-tuning, while significantly reducing memory usage and computational cost—making it particularly suitable for resource-constrained and data-limited applications.

Several studies have demonstrated the feasibility and effectiveness of PEFT in medical imaging tasks. For instance, Baharoon et al. [15] conducted experimental studies using the DINOv2 model for radiological imaging to explore its applicability to medical tasks. Huix et al. [16] systematically investigated the transferability of various large foundation models for medical image classification. Dutt et al. [17] focused on comparing the performance of different PEFT methods across medical imaging datasets. Basu et al. [18] carried out a large-scale, empirically consistent evaluation to comprehensively assess the performance of various PEFT techniques in few-shot image classification tasks.

These studies [15–18] consistently show that in medical image classification settings, PEFT not only substantially reduces the number of trainable parameters and computational cost but also, in certain scenarios, outperforms traditional full fine-tuning. This highlights the high potential and practical value of PEFT. Therefore, this study argues that integrating PEFT into the classification of cytological images of pulmonary metastatic cancer is both feasible and beneficial, particularly for improving the performance of large pre-trained models under few-shot learning constraints.

Given the common characteristics in cellular imagery—namely, high inter-class similarity and large intra-class variability—a growing number of recent studies have applied Fine-Grained Image Classification (FGIC) techniques to few-shot learning tasks. Related findings, the core concept of FGIC is to enhance classification accuracy by mining both global and local features from images or integrating multi-level deep feature representations, thereby distinguishing subcategories that share highly similar visual characteristics.

The design strategies for FGIC can be broadly categorized into two main approaches: (1) mining global and local features from the image, and (2) integrating features across different layers of the model.

For the first approach, Zheng et al. [19] proposed an FGIC method that combines the TinyViT model with object localization (OL) and graph convolutional networks (GCN). This method enhances classification by extracting features from local key regions and modeling their interrelations via GCN. Similarly, Ji et al. [20] adopted a dual-branch Transformer architecture that fuses multi-granular visual information to improve fine-grained visual recognition. Both approaches focus on extracting local discriminative features and fusing them with global context to boost classification accuracy.

For the second approach, Chou et al. [21] proposed the HERBS (High-temperature Refinement and Background Suppression) method, which maintains consistency between low- and high-level features. Wang et al. [22] designed a multi-granular attention sampling module that extracts feature maps from various layers of the network, allowing the model to focus on discriminative regions while integrating multi-level features. These methods emphasize multi-level deep feature fusion and background noise suppression, demonstrating strong performance in distinguishing visually similar categories.

In summary, the methods proposed by Zheng and Ji [19–20] primarily extract local feature maps and combine them with global features for classification. These approaches are generally more flexible and intuitive in design, enabling the model to focus on key image regions and assist in decision-making. Moreover, they can be implemented as modular add-ons compatible with various backbone architectures, with relatively low requirements for architectural adaptation.

In contrast, the strategies adopted by Chou and Wang [21–22] emphasize integrating hierarchical feature representations across the model. By filtering background noise and combining high-level semantic and low-level spatial features, these methods enhance the completeness and precision of classification. However, they rely more heavily on specific architectures—such as CNNs or Swin Transformers—and often have stricter requirements on input image dimensions, making them comparatively less flexible.

Overall, the literature [19–22] confirms that incorporating FGIC into few-shot learning tasks is not only feasible but also effective in enhancing model discriminability for visually similar categories. This further indicates that FGIC techniques hold great potential for application in our study on cytological image classification of pulmonary metastatic cancer. Considering model compatibility and implementation flexibility, this study will prioritize FGIC approaches that focus on combining global and local features as the primary design direction for the initial classification framework.

Although Vision Transformers (ViT) have demonstrated strong capabilities in global feature modeling for image classification tasks and have achieved impressive performance on large-scale datasets, their effectiveness often falls short of convolutional neural networks (CNNs) of comparable size when trained on small to medium-sized datasets. One key reason is that ViTs lack the inductive biases inherent to CNNs for capturing local relationships and spatial structures, making them less sensitive to fine-grained regions and less effective at focusing on image areas critical to classification.

To address this limitation, recent studies [23–26] have proposed hybrid model designs that integrate CNN architectures into ViTs to enhance classification performance under few-shot learning conditions. Wang et al. [23] proposed the Unified Query-Support Transformers (QSFormer), which combines global sample relationship modeling (sampleFormer) and local image structure learning (patchFormer) to improve consistency and metric learning between query and support sets or individual images. Their model also incorporates a Cross-Scale Feature Extractor (CIFE), which utilizes CNNs to extract feature maps at multiple scales for more effective representation learning. Askari et al. [24] introduced a multi-output embedding network with an attention mechanism that channels feature outputs from each model layer into a self-attention module, enhancing the model's ability to capture critical features. Dai et al. [25] proposed the CoaTNet architecture, which merges the strengths of convolution and Transformer modules through relative attention, depthwise convolution, and shortcut connections, preserving features from earlier layers to improve generalization and efficiency. Wu et al. [26] developed the Convolutional Vision Transformer (CvT), which incorporates CNN-like characteristics into ViT to achieve the local invariance of CNNs and the global modeling capability of Transformers, thereby strengthening the model's ability to extract local features.

These findings confirm that integrating CNN components into ViT architectures not only compensates for ViT's weakness in local feature representation but also preserves its advantage in global context modeling. By combining the strengths of both structures, hybrid models significantly improve overall classification performance. For the cellular image classification task in this study, adopting such a hybrid architecture is both feasible and beneficial, as it enhances the model's attention to critical cell regions and improves recognition accuracy.

In summary, the aforementioned methods demonstrate the effectiveness and potential of hybrid CNN–Transformer architectures in few-shot learning scenarios. Given that most of these methods yield strong experimental results, this study adopts the hybrid model strategy, integrating CNN and Transformer strengths. Specifically, the Convolutional Vision Transformer (CvT) is selected as the primary backbone model due to its demonstrated high performance on public datasets, well-structured architectural design, and adaptability to various task requirements. For these reasons, CvT will serve as the core model for the cellular image classification task in this research.

Finally, in few-shot learning tasks, limited training data often leads to overfitting, which negatively impacts the model's generalization ability. To mitigate this issue, an increasing number of recent studies [27–30] have applied Contrastive Learning (CL) to image classification tasks to enhance model performance in few-shot scenarios. The core idea of contrastive learning is to optimize the feature space such that representations of similar samples are pulled closer together, while those of dissimilar samples are pushed further apart. This strategy helps alleviate overfitting caused by data scarcity and improves the model's ability to generalize to unseen classes or domains.

For instance, Cheng et al. [27] proposed the FGFL (Frequency-Guided Few-shot Learning) framework, which leverages frequency-domain information to improve classification performance, overcoming the limitation of traditional methods that focus solely on spatial features. By masking selected frequency ranges and employing multi-level metric learning, the model extracts more discriminative features. Yang et al. [28] introduced a self-supervised contrastive loss that operates on feature vector vs. feature map and feature map vs. feature map pairs, integrating global and local features to enhance model generalization and transferability. Lim et al. [29] advanced this further by proposing Self-supervised Contrastive Learning (SCL), which incorporates multiple self-supervised tasks—such as image-level contrast and rotation prediction—to strengthen feature representations under few-shot constraints. Ouali et al. [30] developed an attention-based spatial contrastive loss, which learns local but class-agnostic features, thereby improving the model's generalization and transfer capability.

The collective findings of these studies [27–30] indicate that well-designed contrastive objectives can significantly improve the quality of learned feature representations, reduce overfitting, and enhance generalization in few-shot learning settings. Moreover, these works commonly apply image-level preprocessing or augmentation before computing contrastive loss, allowing the model to better distinguish between subtle differences. Inspired by these approaches, our study also incorporates a contrastive learning strategy, where preprocessed image pairs are used to calculate task-specific contrastive losses aligned with the classification objective. This serves as a key auxiliary technique for improving classification performance in our cytological image classification task.

Based on the aforementioned literature, it is evident that few-shot learning (FSL) enables effective model training and classification even with a limited amount of data. This approach first employs meta-learning to establish a broad knowledge foundation within the model, followed by fine-tuning to adapt the learned knowledge to specific tasks, thereby enhancing task performance. Furthermore, adopting Parameter-Efficient Fine-Tuning (PEFT) strategies can significantly reduce computational resource consumption while effectively fine-tuning large-scale models with limited data.

In addition, integrating fine-grained image classification (FGIC) techniques allows for more accurate discrimination of subtle yet critical morphological differences between cell categories. To further enhance model performance, this study adopts a hybrid model architecture that incorporates the strength of CNNs in local feature extraction into the ViT framework, and combines it with contrastive learning to improve the model's generalization and feature discriminability, thereby comprehensively boosting classification performance.

However, a review of existing studies reveals a lack of classification models specifically designed for cytological images of pulmonary metastatic cancer. Therefore, this study proposes a deep learning model tailored for few-shot learning scenarios, specifically developed for the classification of pulmonary metastatic cancer cell images. This model aims to provide a feasible and practical solution for clinical applications.

To enable effective feature learning from cytological images of pulmonary metastatic cancer and support clinical decision-making, we propose a novel model tailored to this classification task by adapting and improving existing few-shot learning (FSL) techniques.

Recent advances in FSL have demonstrated the feasibility of training models with limited data. However, there are no publicly available datasets for PMC cytology. To address this, we adopt the Pre-train → Meta-learn → Fine-tune (PMF) framework, which incorporates a fine-tuning stage. Recognizing the limitations of full fine-tuning in low-data regimes, we further incorporate Parameter-Efficient Fine-Tuning (PEFT). In response to the morphological similarity among cell types, we integrate Fine-Grained Image Classification (FGIC) techniques. To compensate for the weak local feature modeling of Vision Transformers (ViTs), we adopt a hybrid architecture that combines CNN and Transformer modules. Finally, to address the overfitting problem inherent in few-shot learning, we introduce contrastive learning strategies. The final architecture is referred to as PFLCH.

The main contributions of this study are as follows:

(1) We design a computer-aided diagnostic system that assists physicians in the preliminary classification of cancer types, thereby reducing diagnostic delay and expediting treatment planning.

(2) We propose a few-shot learning model tailored for PMC datasets, which are characterized by numerous categories, limited samples per class, and the lack of public datasets.

(3) We introduce a fine-grained classification method to improve discriminability in the presence of inter-class morphological similarity among cancer cell types.

(4) To address domain shift and data scarcity, we implement a parameter-efficient fine-tuning strategy to improve adaptability with minimal data.

(5) To compensate for ViT's limitation in local feature extraction, we design a hybrid model that integrates CNN with Transformer architectures.

(6) To mitigate overfitting under low-data regimes, we incorporate a contrastive learning framework tailored to cytological image representations.

In summary, this study proposes a deep learning–based PFLCH computer-aided diagnostic system designed specifically for the classification of pulmonary metastatic cancer cytological images. The system aims to enhance diagnostic performance over existing methods, assist physicians in early detection and treatment planning, reduce mortality and clinical cost, and improve current diagnostic workflows in real-world medical settings.

## 2. Related Work

### 2.1. Few shot learning (FSL)

Few-Shot Learning (FSL) is a machine learning paradigm designed to train models using a limited amount of labeled data, enabling them to maintain strong predictive performance when faced with new tasks. The core idea of FSL is to train the model by randomly generating multiple tasks from a source dataset, so that it learns to generalize under data-scarce conditions.

During each training epoch, a set of tasks is constructed and input into the model for training. The model is then evaluated on a target dataset, which also consists of multiple tasks designed to assess classification performance. As illustrated in Figure 2-1, each task comprises a Support Set and a Query Set. The support set functions similarly to a training set and is used to help the model learn class distinctions, while the query set serves as the testing set, evaluating how well the model can predict unseen examples.

A commonly used terminology in FSL is the "N-way K-shot" setting, which refers to a task consisting of N classes, with K images per class in the support set. The number of query images per class can be set as needed and is used to test the model's recognition accuracy. Through this N-way K-shot task formulation, FSL achieves effective training and accurate prediction even with limited data.

Methods in Few-Shot Learning (FSL) can be broadly categorized into two main approaches: Metric Learning and Model-Agnostic Meta-Learning (MAML).

The core idea of Metric Learning is to learn a feature embedding space in which similar images are mapped closer together while dissimilar images are mapped farther apart. Classification is then performed based on the similarity between image embeddings. Representative approaches include Prototypical Networks [31], Matching Networks, Siamese Networks, and models using specific loss functions such as the Triplet Loss.

On the other hand, MAML is a general-purpose meta-learning framework that learns an initialization of model parameters that can quickly adapt to new tasks with a few gradient updates. Common methods include the original MAML algorithm and its variant Reptile. However, MAML typically requires iterative gradient computation and aggregation across multiple tasks, resulting in higher computational cost.

Considering computational efficiency and practicality, most current few-shot classification methods tend to adopt metric learning–based approaches due to their lower computational overhead [4–6,10,14]. Among these, Prototypical Networks [31] are regarded as a classic and effective solution. This method classifies query images based on the distance between their embeddings and class-specific prototypes (i.e., the mean vector of the support examples in each class). The class with the closest distance is predicted as the label. Compared to other methods, Prototypical Networks offer several advantages: multi-class scalability, conceptual simplicity, computational efficiency, and strong interpretability.

Therefore, this study adopts Prototypical Networks [31] as the foundational model for our experiments, with plans to integrate additional methods in future work to further optimize model performance..

### 2.2. Parameter-efficient Fine-Tuning (PEFT)

Parameter-Efficient Fine-Tuning (PEFT) is a technique designed to reduce the number of parameters required when fine-tuning large pre-trained models. Instead of updating the entire network, PEFT selectively adjusts only a small subset of parameters, while preserving the majority of the original pre-trained weights. This strategy effectively reduces computational resource consumption and improves fine-tuning efficiency, all while maintaining performance close to that of full fine-tuning.

Numerous studies related to few-shot learning have adopted PEFT methods [15–18], particularly within the medical imaging domain [15–17], where PEFT has been widely applied and deeply explored. For example, [15] utilized the powerful DINOv2 pre-trained model and applied PEFT techniques to classify radiological images. The results showed that although the classification performance was comparable to full fine-tuning, the number of parameters updated was significantly reduced, highlighting PEFT's clear advantage in lowering computational costs. Other works [16–18] have reported similar outcomes, indicating that PEFT can achieve equal or even superior performance compared to full fine-tuning, while substantially reducing the number of trainable parameters, thus enabling more efficient training workflows.

PEFT methods can be classified into several strategies, including Adapter Tuning, Prompt Tuning, Prefix Tuning, Side Tuning, Specification Tuning, and Reparameter Tuning, each offering different trade-offs in terms of efficiency and performance.

According to [32], when the goal is to perform low-parameter fine-tuning without modifying the architecture of the pre-trained model and without introducing inference-time latency, two approaches have emerged as commonly used and effective choices in few-shot learning: BitFit (a type of specification tuning) and LoRA (Low-Rank Adaptation, a type of reparameter tuning) [33]. BitFit updates only the bias terms in the model and is one of the most lightweight fine-tuning methods available. However, it may be less effective in complex tasks such as diverse image classification. In contrast, LoRA, though requiring slightly more trainable parameters, has demonstrated superior performance and stability in more complex scenarios, such as medical image classification [15,17], making it a leading PEFT method in current applications..

### 2.3. Fine-grained image classification(FGIC)

Fine-Grained Image Classification (FGIC) is designed to address classification tasks in which subcategories within a broader category exhibit only subtle visual differences. This technique enhances classification precision by leveraging features at different levels or by jointly extracting both local and global features, enabling the model to simultaneously attend to global structures and capture fine-grained regional variations. Through these mechanisms, the model learns to identify subtle distinctions between subcategories and thus achieves effective fine-grained classification.

Many recent studies in the field of few-shot learning have applied FGIC techniques to datasets characterized by minimal inter-class variation, such as bird or aircraft image classification. For example, the method proposed in [19] combines an Object Location module with a Graph Convolutional Network (GCN). It first employs the attention mechanism of the TinyViT model to identify regions of interest, extracting local areas that are then fed into a backbone network for deeper feature representation. Spatial relationships among these regions are then modeled using GCN, and local and global features are subsequently fused to perform the final classification. Experimental results show that this approach outperforms baseline models on tasks involving subtle inter-class differences. Similar outcomes have been reported in other studies [20–22], where the integration of fine-grained features has led to significant improvements in handling visually similar image classes.

In general, FGIC methods follow two main design strategies: one involves fusing local features with global features during prediction, and the other enhances the model's overall representation capability by integrating hierarchical features across different network layers. For instance, in the widely used CUB-200-2011 dataset, the method proposed in [19], which falls under the former category, has demonstrated strong performance in ViT-based architectures. Meanwhile, recent advancements have also optimized Swin Transformer–based architectures. Studies such as [21] and [22] have introduced techniques to suppress background noise across multiple feature layers to improve classification accuracy. In particular, the HERBS (High-temperature Refinement and Background Suppression) method introduced in [21] maintains consistency between low- and high-level features and selects the top-k most predictive features for classification. It is currently regarded as one of the most advanced methods in this area.

However, HERBS is built upon the Swin Transformer architecture and is not compatible with the hybrid model architecture adopted in this study. Consequently, we employ the Object Location module proposed in [19] as our fine-grained feature extraction strategy. This module enables the combination of local features with global representations for classification, which is particularly beneficial in enhancing recognition performance in few-shot learning scenarios involving cytological images with subtle morphological differences between classes.

### 2.4. Hybrid Model

Hybrid models refer to architectural designs that combine two or more distinct types of model structures, aiming to leverage the strengths of each to improve overall classification performance, generalization capability, or computational efficiency. In the field of deep learning for image classification, the most commonly adopted hybrid models are those that integrate Convolutional Neural Networks (CNNs) with Transformers. Studies [23–26] have consistently demonstrated that CNNs are highly effective at extracting local features, while Transformers excel at modeling global semantic relationships. By integrating these two architectures, hybrid models can simultaneously capture fine-grained details and global context, thereby significantly enhancing recognition accuracy. These findings suggest that hybrid models consistently outperform models that rely solely on either CNNs or Transformers.

On the ImageNet-1K dataset, the Convolutional Vision Transformer (CvT) model proposed by Wu et al. [26] is considered one of the current state-of-the-art (SOTA) methods. CvT incorporates CNN structures into the Vision Transformer (ViT) architecture to address ViT's limitations when trained on smaller datasets. The model is characterized by two core design components: a hierarchical Transformer architecture and a convolutional projection mechanism. The former utilizes multi-scale convolution operations to extract features across different levels and spatial resolutions, enabling the integration of both local and global information. The latter replaces the original linear projections used to generate Query (Q), Key (K), and Value (V) matrices with convolutional operations, which enhances the model's ability to capture local spatial context, reduces semantic ambiguity, and improves computational efficiency.

Through these innovations, the CvT model effectively combines the scale-invariance of CNNs with the dynamic attention and global modeling capabilities of Transformers, leading to improved performance and inference efficiency in classification tasks. Additional studies [23–25] further support the advantages of CNN–Transformer hybrid architectures, demonstrating superior performance across a variety of classification problems compared to traditional standalone models.

Therefore, in order to enhance the model's capacity for feature recognition and classification, this study replaces the previously adopted ViT-Small feature extractor with the CvT model [26]. This substitution is expected to further improve overall performance and classification accuracy in the few-shot learning setting.

### 2.5. Contrastive Learning (CL)

Contrastive Learning (CL) is a self-supervised learning approach that aims to learn discriminative feature representations by comparing the similarity and dissimilarity between data samples. In few-shot learning (FSL), where the amount of available training data is limited, models are prone to overfitting due to insufficient supervision. As a result, numerous studies have incorporated contrastive learning into few-shot frameworks to enhance the generalization ability of models and, in turn, improve classification performance.

The fundamental idea behind contrastive learning is to design a training objective that pulls similar samples closer and pushes dissimilar samples farther apart in the feature space, thereby mitigating the challenges posed by limited data. For instance, Yang et al. [28] proposed a contrastive loss function that integrates both global and local feature information, combined with random data augmentation techniques. Their method incorporates both self-supervised and supervised contrastive loss components, effectively improving the model's ability to distinguish between individual samples and intra-class variations. Multiple studies [27–30] have also confirmed that carefully designed contrastive learning objectives and loss functions can significantly enhance both learning quality and model generalization.

Among these, the state-of-the-art (SOTA) method on the Mini-ImageNet dataset is the approach proposed by Cheng et al. [27]. This method improves class discriminability through masking operations and combines sample-level and class-level contrastive strategies, substantially improving the model's performance in few-shot classification and generalization to unseen classes. Other studies [28–30] follow similar strategies: applying preprocessing techniques to the input images, followed by the design of contrastive objectives and loss functions to further enhance the model's feature learning and recognition capabilities.

Although Cheng et al.'s [27] approach demonstrates superior performance, the method proposed by Yang et al. [28], which combines random data augmentation with contrastive loss, offers greater flexibility and modularity. This makes it more suitable for integration into various model architectures. Therefore, this study adopts Yang et al.'s [28] approach by incorporating contrastive learning into the proposed model and applying it to locally extracted patches obtained through fine-grained image classification.

Specifically, we implement a random masking operation on local image patches and design a contrastive loss between the original global image and its masked local variant. This strategy reinforces the model's ability to maintain representation consistency between the two views, thereby improving both discriminative capability and generalization performance in the few-shot classification task.

## 3. Material and Methods

### 3.1. Dataset and Preprocessing

The cytology image dataset used in this study was provided by the National Taiwan University Cancer Center (IRB Approval No.: 202407027RINA) and consists of images collected during endobronchial ultrasound (EBUS) procedures. This work focuses on the classification of pulmonary adenocarcinoma and metastatic cancers originating from the breast and colon, which are among the most commonly observed pulmonary metastases. Accordingly, the test set includes only cytological images from these three cancer types.

For the training and validation phases, we utilized the Mini-ImageNet dataset [34], adopting the official training and validation splits to construct a simulation environment for few-shot learning. A total of 600 training tasks and 150 validation tasks were generated using a 3-way 5-shot configuration. To ensure robust evaluation in the testing phase, we constructed 600 testing tasks with the same 3-way 5-shot setup. Special care was taken to prevent data leakage between the support and query sets by selecting patient cases such that no single individual contributed to both sets within the same task. Specifically, patients were randomly assigned to the query set first, and the support set was then formed by selecting five images per class from non-overlapping cases. Model performance was assessed by averaging classification accuracy across all test tasks.

Data augmentation was applied consistently during both training and fine-tuning stages to improve model generalization. Common augmentation techniques, such as random resized cropping, color jittering, horizontal and vertical flipping, and random affine transformations, were employed. In addition to these standard procedures, two domain-specific augmentation methods were developed based on the characteristics of cytology images. The first, local patch cropping, was designed to help the model focus on subtle morphological differences between cell types by isolating local structures. The second, random masking, was used to simulate the presence of noise or occlusions, such as those caused by imaging artifacts or debris, thereby encouraging the model to maintain classification stability under partially obscured inputs. These augmentations, especially tailored for cytology images, serve to enhance the model's capacity to learn discriminative features in a low-data setting.

### 3.2. Experimental Procedure

The experimental procedure is illustrated in Figure 2. This study begins with data collection. The training and validation datasets are obtained from the publicly available Mini-ImageNet dataset [34], while the testing dataset is constructed based on the classification of different cancer types.

In terms of model design, we propose a few-shot learning model tailored for the classification of pulmonary metastatic cancer cytology images.

Finally, a series of evaluations are conducted, including:

(1) Comparison of the performance of various few-shot learning models,
(2) Analysis of differences across different model architectures,
(3) Evaluation of different fine-tuning strategies,
(4) Ablation studies of various model components,
(5) Testing under different shot settings, and
(6) Analysis of the impact of different training data sources.

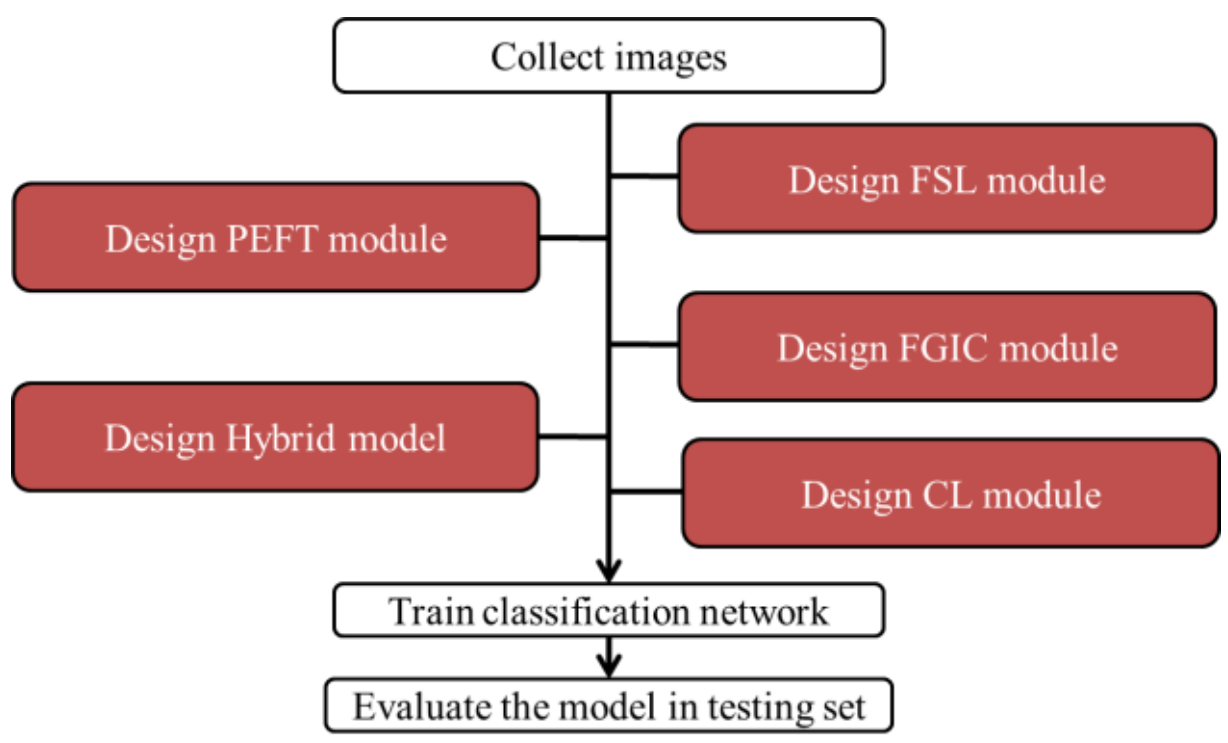

Figure 2. Experiment process

### 3.3. Model Overview

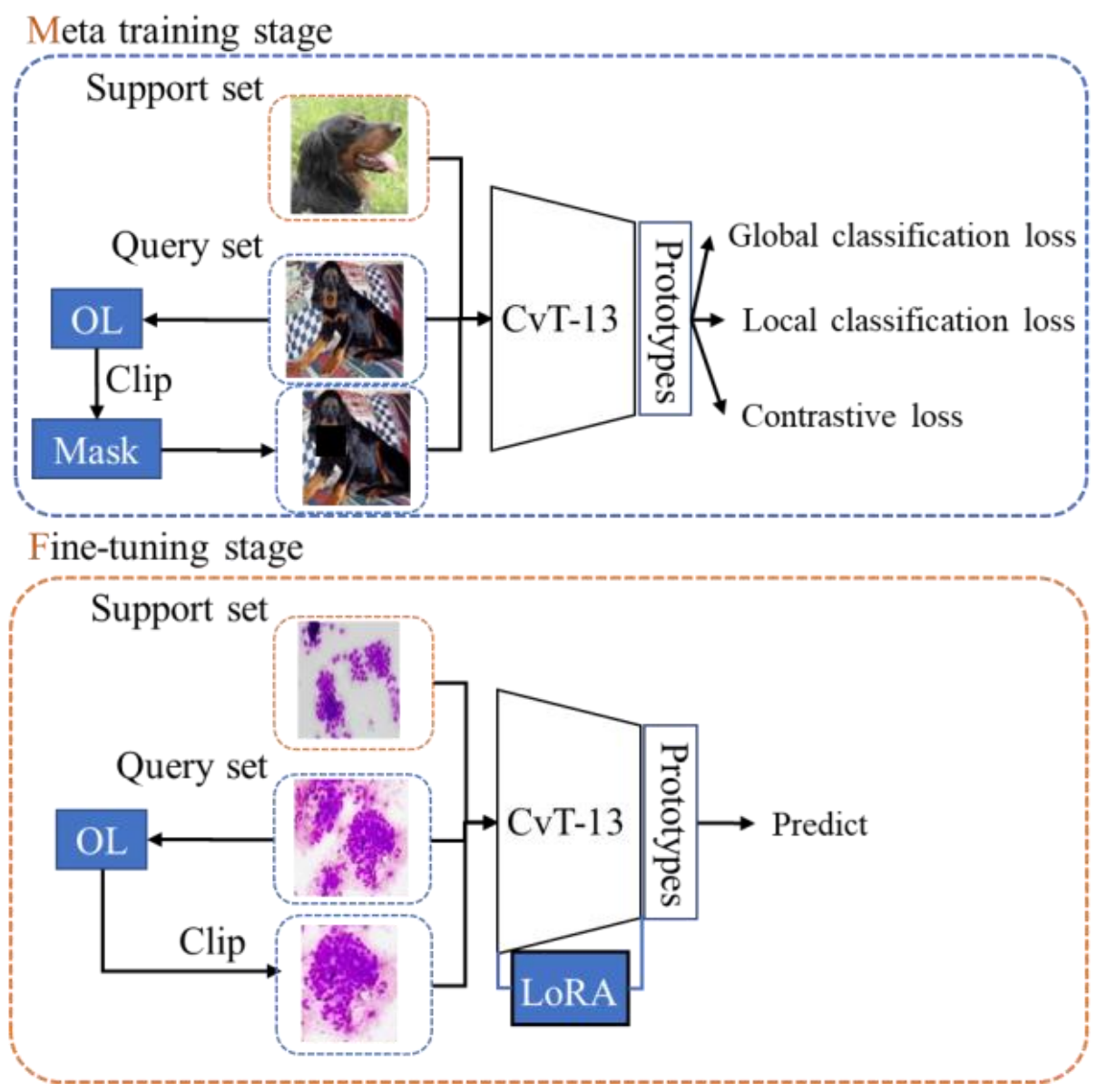

Figure 3. Model overview

The model proposed in this study is referred to as PLFCH (PMF with LoRA, Fine-Grained Classification, Contrastive Learning, and Hybrid Models), as illustrated in Figure 3. PLFCH is built upon the training framework proposed in PMF [10], which consists of a meta-training phase and a fine-tuning phase. The following sections will introduce each phase in detail.

#### *3.3.1. Meta-training stage*

As shown in Figure 4, the meta-training phase primarily aims to train a pre-model with strong generalization capability, enabling it to perform accurate classification even when only a small amount of data is available. We adopt a 3-way 5-shot setting for training, where the model learns from three classes with five samples each in every task. The method used in meta-training belongs to the metric-based category, with Prototype Network [31] being one of the best-performing representatives. The core concept involves inputting the support set images into the model to obtain their feature representations. The feature vectors of the same class are then averaged to establish a class center (prototype). Subsequently, the query set images are fed into the model to obtain their features, and prediction is made by calculating the distance to each class prototype — the closer the distance, the higher the likelihood the image belongs to that class.

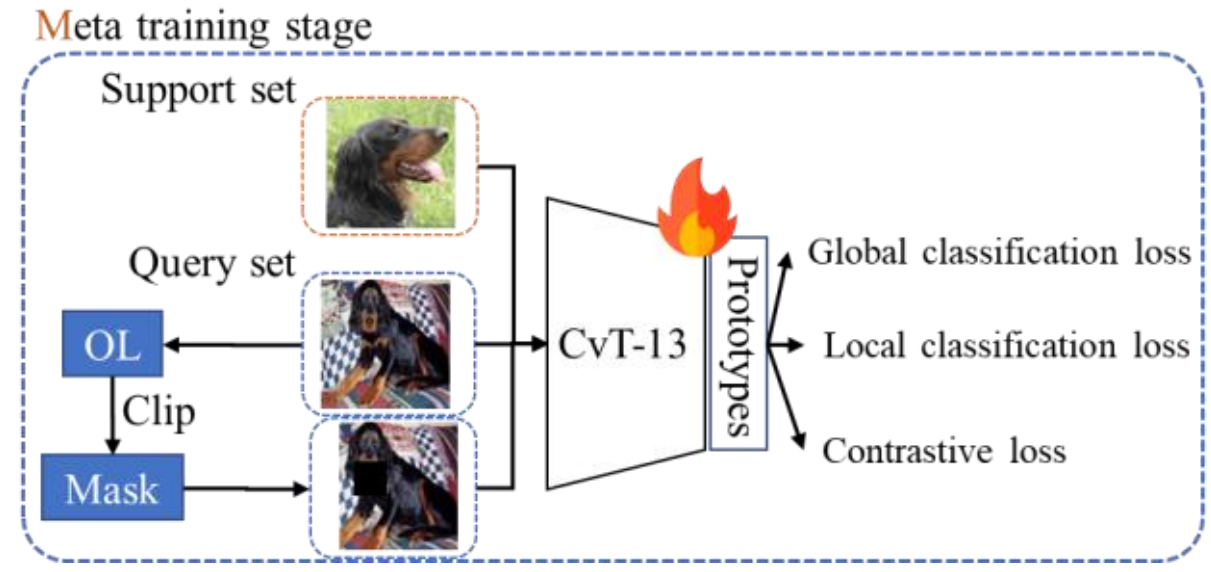

Figure 4. Architecture of the Meta-Training Stage

To further enhance the model's regional recognition ability, we combine the attention weights from the last two layers of the CvT and input them into an object location module to extract the regions the model focuses on most. These localized feature maps are then augmented with random masks to simulate occlusion scenarios, and re-fed into the model for prediction. This introduces a contrastive learning strategy that encourages stable recognition of important regions.

For the loss function design, we integrate three types of loss during training: classification loss for the global image, classification loss for the local image, and contrastive loss between global and local predictions. The first two use cross-entropy loss to optimize the model's classification ability under different perspectives. The contrastive loss uses mean squared error loss (MSE Loss), aiming to measure the difference between the model's predictions on the original full image and the masked local image. By minimizing this difference, we expect the model to maintain consistency in predictions across varying perspectives. The formula for MSE is as follows::

$$MSE(C(X_{Global_Q}), C(X_{Local_Q})) = \frac{1}{n}\sum_{i=1}^{n}\left(C(X_{Global_Q})_i - C(X_{local_Q})_i\right)^2 \quad (1)$$

Here, $C(.)$ denotes the Prototypical classifier, $X_{Global_Q}$ represents the global query image, and $X_{Local_Q}$ represents the local (masked) query image. The variable n denotes the number of query images. The overall loss is defined as follows:

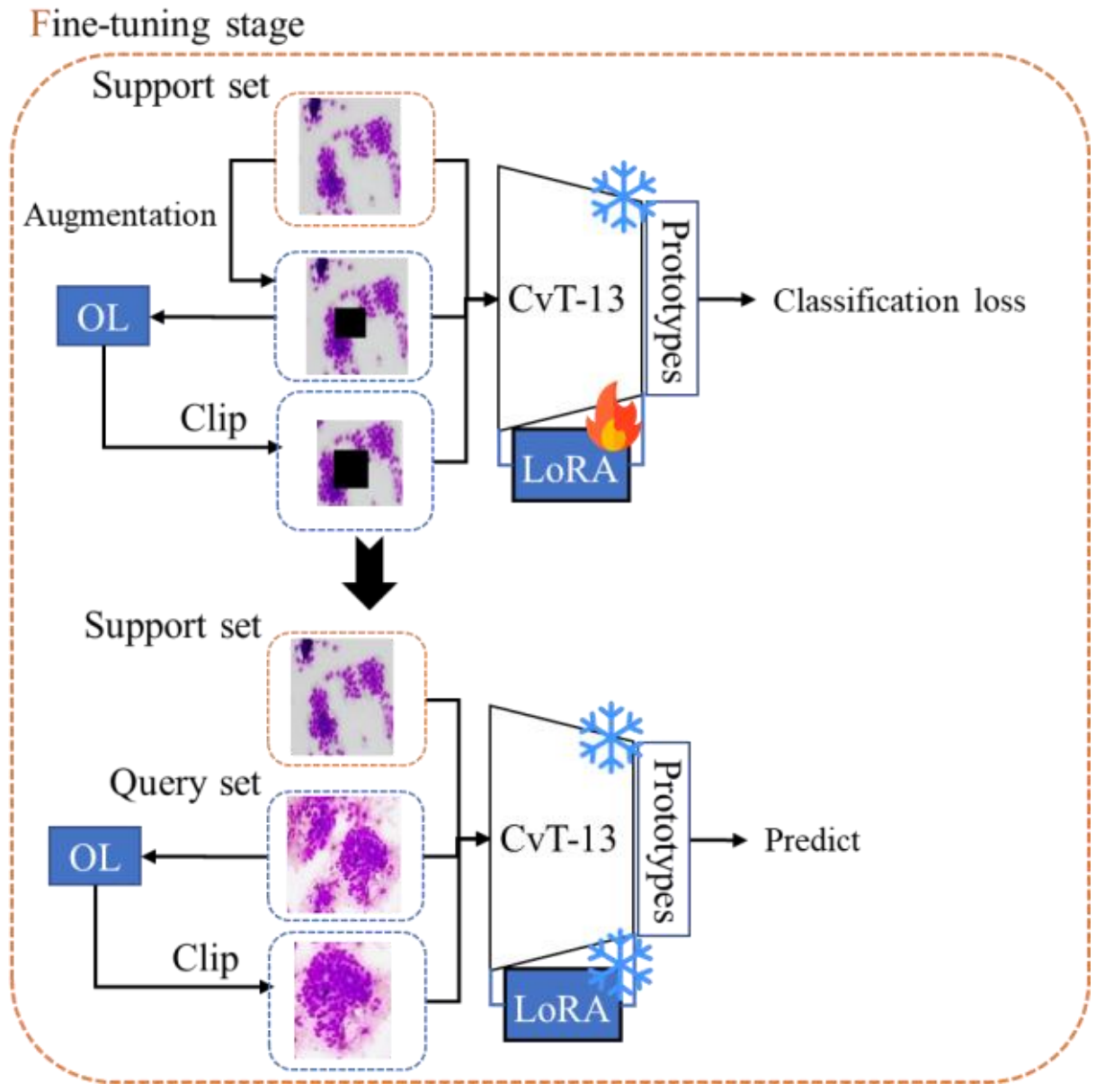

Figure 5. Architecture of the Fine-Tuning Stage

$$\mathcal{L}_{total} = \ell_{cls}\left(C\left(X_{Global_Q}\right), Y_Q\right) + \ell_{cls}\left(C\left(X_{Local_Q}\right), Y_Q\right) + MSE\left(C\left(X_{Global_Q}\right), C\left(X_{Local_Q}\right)\right) \quad (2)$$

Here, $\ell_{cls}$ denotes the classification loss, which is computed using cross-entropy between the model's predictions (either from the global or local query images) and the ground truth labels. $C(.)$ represents the Prototypical classifier, $X_{Global_Q}$ and $X_{Local_Q}$ denote the global and local (masked) query images respectively, and $Y_Q$ denotes the corresponding ground truth labels. The model weights that achieve the best performance on the validation set during this phase will be selected for subsequent fine-tuning.

*3.3.2. Fine-tuning stage*

To adapt the model to our cytology image dataset, this phase builds upon the previous meta-trained model, which has acquired broad knowledge. As illustrated in Figure 3-5, transfer learning is employed to transfer the model's prior knowledge to the specific task of this study. Since previous studies have shown that fine-tuning an entire model with only a small amount of data yields limited results, we employ a parameter-efficient fine-tuning method, LoRA [33], to optimize the parameters of the Attention layers in the CvT architecture.

The fine-tuning process is as follows: we first perform data augmentation on the support set to generate a pseudo query set. This pseudo query set is processed by the object location module to produce localized images, which are then fed into the model to extract features. The support set is used to construct a Prototypes classifier, which is subsequently used for predicting both global and local versions of the pseudo query set. Finally, the global and local prediction results are summed and compared with the labels to compute classification loss for fine-tuning. After fine-tuning is completed, the actual query set is input into the model, and the final classification result is obtained by combining global and local predictions.

### 3.4. Equipment

The experiment uses an ASUS Z790-A GAMING WIFI 6E server equipped with an Intel i9-13900K processor and a single MSI RTX 4090 GAMING X TRIO 24G graphics card.

## 4. Result

### 4.1. Evaluation Methods

For the classification model, this study uses accuracy, precision, recall, and F1-score as evaluation metrics. Precision and Recall are used to represent the model's performance in correctly or incorrectly predicting each class, and F1-score is used for overall evaluation. The accuracy, precision, recall, and F1-score presented below are the average results from 600 queries.

Due to the absence of other publicly available datasets related to metastatic cancer, this study only uses the data provided by National Taiwan University Hospital for testing.

### 4.2. Implement Detail

The details of the proposed model are as follows:

Meta-training Phase:

During training, the model's hyperparameters include an initial learning rate of 5e-5, and the pre-trained weights are based on ImageNet1k. A total of 600 tasks were generated in a 3-way 5-shot setting from the training set for training, and 150 tasks from the validation set were used for validation to obtain well-performing model weights. The optimizer used was SGD, and the loss functions were cross-entropy loss and mean squared error loss. The input image size was 224×224. Data augmentation methods applied to the training set included random resized cropping, color jittering, random horizontal flip, random vertical flip, random affine transformation, and random masking. For the validation set, images were resized for input.

Fine-tuning Phase:

For the medical image testing set, 600 tasks were generated under the 3-way 5-shot setting. During fine-tuning, each task was first trained on the support set for 40 epochs with an initial learning rate of 0.001. The fine-tuned model was then used to predict on the query set. The optimizer used was AdamW, and the loss function was cross-entropy loss. The input image size was 224×224. Data augmentation applied to the testing set during fine-tuning included random translation, random brightness, random saturation, random contrast, and random masking.

### 4.3. Comparison of Different Backbones

This section investigates whether using different backbone architectures affects the model's performance. As shown in Table 1, all models were trained using the same pretrained weights from ImageNet-1k, under identical training and validation settings with full fine-tuning. The results reflect the performance of each model on the testing set. It can be observed that the CVT model achieved better classification results, indicating that combining the strengths of CNNs and Transformers can enhance classification performance.

We also observed that ResNet-18, with significantly fewer parameters, outperformed ResNet-50 and ViT-Small in terms of classification. This suggests that, under limited data conditions, fine-tuning larger models may be less effective. Although ResNet-18 slightly underperformed compared to ViT-Base, the latter's global attention capability allows it to better distinguish subtle differences in cellular images, enabling it to retain more fine-grained information and ultimately achieve superior performance.

\

Table 1. Comparison of Different Backbones

| **Backbone (Parameters)** | **Accuracy (%)** | **Average-Precision** | **Average-Recall** | **Average-F1 score** |
|---|---|---|---|---|
| **Resnet18 (11.7M)** | 42.05% | 0.410 | 0.421 | 0.415 |
| **Resnet50 (25.6M)** | 41.83% | 0.403 | 0.418 | 0.410 |
| **ViT-small (22M)** | 38.46% | 0.386 | 0.385 | 0.386 |
| **ViT-base (85M)** | 42.57% | 0.401 | 0.426 | 0.413 |
| **CvT (28M)** | **47.04%** | **0.463** | **0.470** | **0.466** |

### 4.4. Comparison of Different Fine-tuning Methods

This section discusses the impact of different fine-tuning strategies on model performance. The results are presented in Table 2. All experiments were conducted using the same CvT model architecture and identical training and validation procedures. In the table, Fine-tuning refers to full model fine-tuning, while Fine-tuning$^{+}$ refers to fine-tuning only the final layer of the model.The results show that LoRA [33], a parameter-efficient fine-tuning method, achieved the best performance with significantly fewer trainable parameters, demonstrating effective adaptation. This is because LoRA [33] optimizes the Attention layers within the CvT architecture using a small number of additional parameters, allowing the model to leverage existing prior knowledge while adapting to our dataset.On the other hand, BitFit, another parameter-efficient approach which fine-tunes only the bias terms, required the least parameter updates but yielded suboptimal results. This is likely due to its limited capacity to adapt to the diverse nature of cytological images. Meanwhile, fully fine-tuning the model or fine-tuning only the final layer involved a larger number of parameters, and such approaches showed limited effectiveness when applied to large models under low-data conditions.

Table 2. Comparison of Different Fine-tuning Methods

| Method | Trainable Params. (%) | Accuracy (%) | Average-Precision | Average-Recall | Average-F1 score |
|---|---|---|---|---|---|
| No fine-tuning | - | 44.66% | 0.435 | 0.444 | 0.439 |
| Fine-tuning | 19.61M(100%) | 47.04% | 0.463 | 0.470 | 0.466 |
| Fine-tuning* | 1.79M(9.11%) | 45.21% | 0.441 | 0.452 | 0.447 |
| BitFit | **0.06M(0.31%)** | 47.40% | 0.471 | 0.474 | 0.472 |
| LoRA | 0.15M(0.74%) | **47.93%** | **0.474** | **0.479** | **0.477** |

### 4.5. Ablation Study

This section presents an ablation study to analyze the contribution of each component in the proposed method, as shown in Table 3. The method integrates several key strategies: a hybrid model (CvT) is employed as the feature extractor, replacing the original ViT-Small to enhance image representation capability. Parameter-efficient fine-tuning (PEFT) using LoRA is applied to adapt the CvT model to the dataset with limited data. Fine-grained image classification (FGIC) is introduced by incorporating an Object Location (OL) module to extract local features. Lastly, a contrastive learning (CL) approach is adopted, which applies random masking to local image regions and employs a mean squared error (MSE) loss to enforce consistency between masked and original representations.

The results demonstrate that each of these components improves classification performance. Notably, the hybrid model and contrastive learning yield the most significant gains, improving accuracy by 3.04% and 1.59% respectively, thereby enhancing both the classification performance and the generalization ability of the model. We speculate that the use of the CvT hybrid model, which combines the advantages of CNNs and Vision Transformers, significantly enhances the model's ability to extract and discriminate visual features compared to the original ViT, resulting in improved classification outcomes. In the case of contrastive learning, the use of random masking introduces information loss in the local image regions, increasing the difficulty of the task. The MSE loss compels the model to maintain consistent predictions between the original and the masked images, which strengthens its ability to recognize discriminative features and improves generalization.

Although the improvements from PEFT and FGIC are relatively modest, the PEFT method reduces the number of tunable parameters, enabling large models to be effectively fine-tuned even under limited-data conditions while also reducing memory usage and computational load. FGIC, on the other hand, allows the model to capture fine-grained visual details by focusing on local regions, which, when combined with contrastive loss, further amplifies the classification performance.

With all four methods combined, the final model achieves an accuracy of 49.59%, representing a 5.59% improvement over the baseline. Other metrics also show noticeable gains: precision increased by 0.039 to 0.479, recall by 0.056 to 0.496, and F1-score by 0.048 to 0.488. These results suggest that the proposed framework is specifically tailored to the characteristics of the lung metastasis dataset and effectively improves classification performance.

Table 3. Ablation Study

| Hybrid model | PEFT | FGIC | CL | Accuracy (%) | Average - Precision | Average - Recall | Average - F1 score |
|---|---|---|---|---|---|---|---|
| | | | | 44.00% | 0.440 | 0.440 | 0.440 |
| ✓ | | | | 47.04% (+3.04%) | 0.463 | 0.470 | 0.466 |
| ✓ | ✓ | | | 47.93% (+3.93%) | 0.474 | 0.479 | 0.477 |
| ✓ | ✓ | ✓ | | 48.00% (+4.00%) | 0.462 | 0.480 | 0.471 |
| ✓ | ✓ | ✓ | ✓ | **49.59% (+5.59%)** | **0.479** | **0.496** | **0.488** |

### 4.6. Comparison of Models

This section aims to compare the proposed PLFCH method with other few-shot learning (FSL) classification models. All methods were trained and validated using the same dataset, Mini-ImageNet [34], with each model trained using its recommended hyperparameters and architecture. Finally, all models were evaluated on our custom dataset for testing. Table 4 presents the results of various few-shot learning models on the testing set.

The results clearly indicate that the proposed PLFCH method outperformed all other methods, achieving the highest accuracy of 49.59%, as well as the best precision, recall, and F1-score of 0.479, 0.496, and 0.488, respectively. These outcomes demonstrate that the PLFCH approach is specifically tailored to the characteristics of the lung metastasis dataset and is thus more effective in this application context.

In comparison with more recent state-of-the-art methods such as CAML [11] and PMF [10], CAML emphasizes context-aware learning that enables testing without the need for fine-tuning. However, the domain gap between the Mini-ImageNet dataset and our cytology image dataset remains substantial, resulting in poor classification performance. Additionally, CAML uses pretrained CLIP weights, which, as noted in its original paper, tend to group morphologically similar patterns into the same category. This characteristic may not align well with the nuanced features of our dataset, contributing to the unsatisfactory results.

As for the PMF [10] method, which performs fine-tuning of large models on small datasets, its performance was still inferior to our approach using parameter-efficient fine-tuning via LoRA [33]. This suggests that parameter-efficient strategies like LoRA offer distinct advantages in few-shot learning scenarios, enabling more effective model adaptation with limited data.

These findings, as summarized in Table 4, affirm the effectiveness of our PLFCH method in addressing the specific challenges posed by few-shot classification tasks in the context of lung metastasis cytology images.

Table 4. Comparison of Models methods

| Method | Accuracy(%) | Average-Precision | Average-Recall | Average-F1 score |
|---|---|---|---|---|
| ProtoNet [31] | 33.73% | 0.337 | 0.337 | 0.337 |
| MetaOpt[35] | 34.94% | 0.351 | 0.349 | 0.350 |
| LDPnet [14] | 36.75% | 0.373 | 0.373 | 0.373 |
| PMF [11] | 41.72% | 0.417 | 0.417 | 0.417 |
| CAML [12] | 32.95% | 0.319 | 0.330 | 0.324 |
| PLFCH(Ours) | **49.59%** | **0.479** | **0.496** | **0.488** |

**4.7 Evaluation with Varying Numbers of Shots**

This section further investigates the training and testing performance under different numbers of shots. The results are presented in Table 5. Under consistent training strategies and fine-tuning procedures, only the number of shots used for training and testing was varied.

The experimental results show a clear trend: as the number of shots increases, classification performance improves accordingly. Specifically, the accuracy increased from 44.84% with 1-shot to 49.59% with 5-shot—the setting used in this study. Further increases to 10-shot and 20-shot yielded accuracies of 54.62% and 55.48%, respectively, with the 20-shot configuration achieving the best performance.

In the context of few-shot learning, the number of shots represents the number of support images available for the model to refer to during training and prediction. More support images provide richer contextual cues, enabling the model to better distinguish between classes and thus improving classification accuracy. This effect is especially significant for our medical cytology dataset, which exhibits small inter-class differences and large intra-class variability. Providing more support images (e.g., 10 or 20 per class) allows the model to learn more stable and representative features, enhancing its ability to differentiate between visually similar classes.

Table 5. Evaluation with Varying Numbers of Shots

| Shots | Accuracy(%) | Average-Precision | Average-Recall | Average-F1 score |
|---|---|---|---|---|
| 1 | 44.84% | 0.450 | 0.449 | 0.449 |
| 5 | 49.59% | 0.479 | 0.496 | 0.488 |
| 10 | 54.62% | 0.536 | 0.546 | 0.541 |
| 20 | 55.48% | 0.556 | 0.555 | 0.555 |

**4.8 Impact of Different Training Data Sources**

This section investigates the effect of using different training data sources on model performance. The results are summarized in Table 6. Under consistent training strategies and fine-tuning procedures, the only variable altered was the choice of training dataset. In this study, the primary training dataset used was the publicly available Mini-ImageNet [34], which consists of natural scene images encompassing a wide range of object categories such as animals, tools, food, and vehicles. This dataset is commonly used in few-shot learning research, with 64 classes in the training set and 16 in the validation set.

To explore whether using domain-relevant medical imaging datasets could reduce the domain shift between the training and testing data and potentially improve performance, we introduced an alternative training dataset. This dataset was constructed by combining two medical imaging sources: a peripheral blood cell image dataset [36] and the LC25000 dataset [37], which contains histopathological images of lung and colon tissues. The combined dataset was divided into 8 classes for training and 5 classes for validation, and models were retrained and tested on our custom dataset accordingly.

Contrary to our expectations, the results revealed that models trained on the Mini-ImageNet dataset performed better than those trained on the medical imaging dataset. Although we initially hypothesized that training with domain-relevant data would mitigate domain shift and improve fine-tuning performance on our dataset, the findings suggest otherwise.

We interpret this outcome as a result of the greater diversity and richness of categories in Mini-ImageNet, which likely enabled the model to learn more robust and transferable feature representations. Despite being closer in domain, the medical image dataset contained fewer classes and limited variability, which may have restricted the model's generalization ability. Consequently, models trained on Mini-ImageNet demonstrated superior performance when evaluated on our dataset.

Table 6. Impact of Different Training Data Sources

| Method | Accuracy(%) | Average-Precision | Average-Recall | Average-F1 score |
|---|---|---|---|---|
| Mini-imagenet [34] | **49.59%** | **0.479** | **0.496** | **0.488** |
| Medical Imaging[36-37] | 47.05% | 0.450 | 0.470 | 0.460 |

### 4.9. Experiment summary

The study systematically examined various aspects influencing model performance. In Section 4.3, a comparison of different backbone architectures was conducted. Section 4.4 explored the effects of various fine-tuning strategies. Section 4.5 presented an ablation study, analyzing the contribution of each proposed component. Section 4.6 demonstrated that the proposed PLFCH method achieved the highest performance on the testing set, with an accuracy of 49.59%, a precision of 0.479, a recall of 0.496, and an F1-score of 0.488—confirming that PLFCH is a well-designed model tailored to lung metastasis cytology images. Subsequent analyses extended the investigation: Section 4.7 evaluated the impact of different numbers of shots, while Section 4.8 examined the influence of training data sources on model performance.

## 5. Conclusion

In the classification of lung metastasis cytology images, the proposed PLFCH model achieved an accuracy of 49.59% on the testing set, with a precision of 0.479, a recall of 0.493, and an F1-score of 0.488. PLFCH integrates several strategies, including few-shot learning, parameter-efficient fine-tuning, fine-grained image classification, a hybrid model architecture, and contrastive learning. These combined approaches enabled the model to outperform other few-shot classification methods in the specific task of classifying lung metastasis cytology images. Our findings confirm that, in the context of medical imaging, providing more image samples helps the model learn features more consistently, thereby improving its ability to distinguish between visually similar classes. Additionally, we observed that using datasets with a greater number of categories—even if from a different domain—can enhance the model's feature representation capabilities more effectively than training with domain-specific but limited-category datasets.

There is still considerable room for improvement in the current model. Future work may involve validating the model's performance on more established public benchmarks such as Mini-ImageNet. Moreover, with the recent rise of large language models (LLMs), many few-shot learning approaches have begun to incorporate text labels from LLMs to enhance model discrimination. Future designs could leverage LLMs to integrate semantic label information, potentially improving classification performance further. Expanding the range of cell categories included in the dataset could also enhance the model's applicability in clinical settings. However, this study has certain limitations. The current dataset for lung metastasis originates from a single medical institution and contains relatively few samples. Future efforts should focus on acquiring a larger and more diverse dataset, including additional subtypes of lung metastasis, to improve the model's robustness, generalizability, and practical applicability in medical diagnosis.